\documentstyle[11pt,newpasp,twoside,epsf]{article}
\def\edcomment#1{\iffalse\marginpar{\raggedright\sl#1\/}\else\relax\fi}
\reversemarginpar

\begin{document}
\title{Millisecond Radio Pulsars in 47~Tucanae}
\author{D.~R.~Lorimer$^1$, F.~Camilo$^2$, P.~Freire$^3$, M.~Kramer$^1$, A.~G.~Lyne$^1$, R.~N.~Manchester$^4$ \& N.~D'Amico$^5$}
\affil{$^1$U. of Manchester, Jodrell Bank Observatory, Cheshire,
SK11~9DL, UK}
\affil{$^2$Columbia University, 550 W 120$^{\rm th}$ St., New York, NY10027, USA}
\affil{$^3$Arecibo Observatory, HC3 Box 53995, Arecibo, PR 00612, USA}
\affil{$^4$ATNF, CSIRO, P.O.~Box~76, Epping NSW~1710, Australia}
\affil{$^5$Osservatorio Astronomico di Cagliari, 09012~Capoterra, Italy}

\begin{abstract}
We review the properties of the 22 millisecond radio pulsars currently
known in the globular cluster 47~Tucanae, and their implications for
the mass and gas content of the cluster. Further details can
be found in the publications from this project to date (Camilo et
al.~2000; Freire 2000; Freire et al.~2001a,b; Freire et al.~2003). 
Throughout the review, we look ahead to the future results anticipated
from this fascinating cluster.
\end{abstract}

\section{Parkes Observations of 47~Tucanae: 1990--Present}

The globular cluster 47~Tucanae (hereafter 47~Tuc) contains the
highest number of radio pulsars currently known in any cluster, and
about a third of the total number of known cluster pulsars. Early
searches at 50 and 70-cm wavelengths using the Parkes telescope
discovered the first 11 millisecond pulsars in 47~Tuc by the mid 1990s
(Manchester et al.~1990; Manchester et al.~1991; Robinson et
al.~1995). Four of the pulsars had binary companions with a median
orbital period of 30 hr. Significant modulation of the pulsar signals
by interstellar scintillation meant that most pulsars were not
detected regularly at these wavelengths; consequently, timing
solutions were only possible for two pulsars.

Interest in 47~Tuc was renewed in the late 1990s following the
installation of the sensitive 20-cm Parkes multibeam receiver. Using
the central beam of this system, a further nine pulsars (all members
of binary systems) were discovered by Camilo et al.~(2000). The high
incidence of binary systems was largely a result of the use of
acceleration search techniques in this survey which permitted the
detection of short orbital period systems.  In addition to better
sensitivity, regular observations with the 20-cm system provided more
frequent detections of the pulsars. This allowed timing solutions for
16 pulsars to date (Freire et al.~2001a; 2003).  Since 1999, data have
been acquired using a high-resolution ($512\times0.5$ MHz) filterbank
which has resulted in a threefold increase in time resolution (Freire
et al.~2003). Searches of these data are on-going, with recent (so far
unpublished) discoveries of 4.771 and 2.196 ms binary
pulsars. Currently the total number of millisecond pulsars known in
47~Tuc stands at 22.

\section{Profiles, Luminosities and Spin Periods}

The current sample of 22 millisecond pulsars in 47~Tuc display similar
emission properties to their counterparts in the Galactic disk. In
their compilation of pulse profiles, Camilo et al.~(2000) noticed a
similar number of components and incidence of interpulses to the
sample of disk millisecond pulsars studied by Kramer et al.~(1998). A
large sample of pulsars at a common distance means that the flux
density distribution is a direct measure of the luminosity
distribution.  Freire (2000) found that the form of the 1400-MHz
luminosity function over the interval 1--10 mJy kpc$^2$ is a power law
with a slope of --1, similar to that found for the pulsars in M15
(Anderson 1992), and the population of normal and millisecond pulsars
in the Galactic disk (e.g.~Lyne et al.~1998).  Assuming a spectral
index of --2, the observed 1400-MHz luminosities scale roughly to a
400-MHz luminosity interval of 10--100 mJy kpc$^2$.  Based on the
detection of $\sim 20$ pulsars in 47~Tuc, the population of objects
with 400-MHz luminosities above 1 mJy kpc$^2$ beamed towards us is of
order 200 (Camilo et al.~2000).

\begin{figure}[hbt]
\plotone{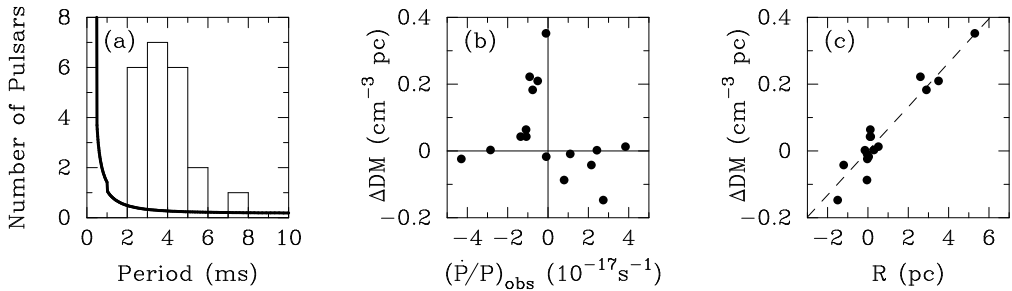}
\caption{(a): Spin period distribution for the 22 currently known
47~Tuc pulsars. The thick line shows the sensitivity curve for the
Camilo et al.~search (linear scale). (b): Dispersion measure relative
to cluster center ($\Delta$DM) versus $\dot{P}/P$ for the 16 pulsars
with timing solutions.  (c): $\Delta$DM versus derived radial position
in the cluster. The dashed line is the best fit assuming a constant
electron density.}
\end{figure}

The spin periods of all 22 pulsars lie in the range 2--8 ms (Fig.~1a).
The absence of long-period pulsars is a real effect.  Particularly
striking is the dearth of pulsars in the 1--2 ms bin compared to 19
objects currently known between 2 and 5 ms. Whether this dropoff at
500 Hz is a real effect (Bildsten, these proceedings) or due to
observational selection is currently a matter for debate.  The
theoretical period sensitivity curve shown in Fig.~1a suggests that
the 1--2 ms pulsars should be almost as easy to detect as the 2--8 ms
pulsars. We are currently searching the high-resolution data to place
much more stringent limits on the pulsar population with periods below
2 ms than the Camilo et al.~search.

\section{Binary Pulsars}

The current population of binary pulsars in 47~Tuc bifurcates into two
main groups: those with orbital periods of order 0.4--2.3 days and
companion masses $\sim 0.2$ M$_{\odot}$ and the so-called very
low-mass binary systems which are characterized by shorter orbital
periods (1.5--5.5 hr) and lighter companions ($\sim 0.02$
M$_{\odot}$). Five binaries (J, O, R, V and W) are eclipsed for some
portion of the orbit by their companion stars.  Of these, J, O and R
belong to the very-low-mass group.

The acceleration searches employed by Camilo et al.~revealed a much
higher incidence of binary systems in 47~Tuc than the earlier
searches. Currently, 15 of the 22 pulsars (68\%) are in binary systems
and the median orbital period is 5 hours.  The shortest orbital period
found so far is the 95-min binary pulsar 47~Tuc~R. This is the
shortest orbital period currently known for any radio pulsar
binary. Could this system and the 11-min orbit of the X-ray source in
NGC~6624 (Stella, Priedhorsky \& White 1987) be the tip of the iceberg
of a large population of short-period binaries in globular clusters?
Population syntheses (Rasio, Pfahl \& Rappaport 2000) suggest that
this may be the case.  Sensitive accleration searches are currently
underway to probe this proposed population.

\section{Astrometry}

Phase-coherent timing solutions currently exist for 16 of the 22
pulsars (Freire et al.~2003) resulting in milliarcsecond positional
determinations (or better in some cases).  All 16 pulsars lie within
1.2 arcmin (4 core radii) of the cluster center, in spite of the fact
that the radius of the 20-cm Parkes beam is 7 arcmin. This
concentration suggests that the pulsars have reached thermal
equilibrium (Rasio 2000; Freire et al.~2001a).  The accurate positions
have enabled {\em CHANDRA} and {\em HST} follow-up work on some of the
pulsars (Heinke et al., these proceedings).  The pulsar radial
distribution is consistent with that of the soft X-ray sources.


Proper motions have now been measured for 11 of the pulsars and upper
limits for 5 others (Freire et al.~2003).  Currently the weighted mean
of the pulsar proper motions is consistent with the optical proper
motion (Odenkirchen et al.~1997) at the 3$\sigma$ level. Pulsar proper
motions are currently dominated by the bulk motion of 47~Tuc. In
future, as the time baseline extends, it should be possible to measure
pulsar motions with respect to the cluster center.

\section{Probing the Mass and Gas in 47~Tuc}

Currently, 10 out of the 16 pulsars with phase-coherent timing
solutions are observed to have negative period derivatives
($\dot{P}_{\rm obs}<0$). Rather than being intrinsic to the pulsars,
the most natural explanation for this apparent spin-up is the
line-of-sight accelerations as the pulsars move within the
gravitational potential of the cluster. Freire et al.~(2001a)
demonstrated that a simple King model potential was consistent with
the observed period derivatives. Neglecting Galactic and proper motion
terms we have $(\dot{P}/P)_{\rm obs} = (a_l/c) + (\dot{P}/P)_{\rm
int}$, where $a_l$ is the line-of-sight acceleration. It follows that
{\it all pulsars with} $\dot{P}_{\rm obs}<0$ {\it are on the ``far
side'' of the cluster.}  Assuming the intrinsic period derivative
$\dot{P}_{\rm int} > 0$ implies $a_l/c > (\dot{P}/P)_{\rm obs}$. A
lower bound on $a_l$ can be used to place a lower bound on the surface
mass density of the matter interior to the pulsar, $\Sigma$ (see
e.g.~Phinney 1992). The most stringent constraint so far is for 47~Tuc S 
(Freire et al.~2003) which lies, in projection, about $12''$ from the 
center of the cluster. For this pulsar $a_l > 1.3 \times 10^{-6}$ cm 
s$^{-2}$ implies $\Sigma > 8.4 \times 10^{4}$ M$_{\odot}$ pc$^{-2}$.

Perhaps the most striking result from the radio pulsars in 47~Tuc to
date is the combination of the above $\dot{P}$ data with
high-precision measurements of the pulsar dispersion measures which
has permitted the detection of ionized gas within the cluster (Freire
et al.~2001b). This is shown in Fig.~1b where those pulsars with
higher dispersion measure are all on the far side of the cluster
($\dot{P}_{\rm obs}<0$).  Under the assumption of a King model
potential, and an intrinsic $\dot{P}/P$ for each pulsar similar to
those known in the Galactic disk, Freire et al.~(2001b) calculated the
radial distance along the line of sight for each pulsar and, as shown
in Fig.~1c, showed that this strongly correlates with dispersion
measure, implying a mean free electron density of $0.067\pm0.015$
cm$^{-3}$. Within the central region of 47~Tuc occupied by these
pulsars, this corresponds to a total gas content of $\sim 0.1$
M$_{\odot}$. This value is much less than the $\sim 100$ M$_{\odot}$
expected to accumulate within the cluster core over $10^{7-8}$ yr
(Roberts 1996).

So where has all the gas gone? As proposed by Spergel (1991), one
mechanism is from the pulsars themselves.  For a typical millisecond
pulsar spin-down luminosity $\dot{E}=10^{34}$ ergs s$^{-1}$, the
energy required to expel the gas can be provided by only 0.5\% of
$\dot{E}$ for a total population of $\sim 200$ pulsars. Somewhat
ironically, the very objects responsible for the detection of gas in
47~Tuc might also be responsible for much of its ejection during the
last billion years.

\acknowledgements
We thank numerous observers who have helped collect 47~Tuc data at
Parkes over the years. DRL is a University Research Fellow funded by
the Royal Society. FC is supported by NASA grant NAG 5-9950.

\end{document}